\documentclass[%
 reprint,
unsortedaddress,
nofootinbib,
 amsmath,amssymb,
 aps,
floatfix,
]{revtex4-2}

\usepackage{amsmath}
\usepackage{euler}
\usepackage{comment}
\usepackage{graphicx}
\usepackage{dcolumn}
\usepackage{bm}
\usepackage[dvipsnames]{xcolor}
\usepackage{braket}

\newcommand{\myexp}{\mathsf{e}}

\begin{document}

\preprint{APS/123-QED}

\title{Order in disordered packings with and without permutation symmetry 
}

\author{Varda F. Hagh}
\email{hagh@illinois.edu}
\affiliation{%
 Department of Mechanical Science and Engineering \\ 
 University of Illinois Urbana-Champaign, 
 Urbana,  IL  61801,  USA.\\}
\author{Sidney R. Nagel}%
\affiliation{%
 Department of Physics and The James Franck and Enrico Fermi Institutes \\ 
 The University  of  Chicago, 
 Chicago,  IL  60637,  USA.\\
}%

\date{\today}
 
\begin{abstract} 
A disordered solid, such as an athermal jammed packing of soft spheres, exists in a rugged potential-energy landscape in which there are a myriad of stable configurations that defy easy enumeration and characterization. Nevertheless, in three-dimensional monodisperse particle packings, we demonstrate an astonishing regularity in the distribution of basin volumes. The probability of landing randomly in a basin is proportional to its volume.  Ordering the basins according to their probability, $P(n)$, from the largest  at $n=1$ to smaller at larger $n$, we find approximately: $P(n) \propto n^{-1}$. This order, persisting up to the largest systems for which we can collect sufficient data, has implications for the  dynamics of a system as it evolves under perturbations. In monodisperse packings there is ``permutation symmetry'' since identical particles can always be interchanged without affecting the system or its properties. Introducing any distribution of radii breaks this symmetry and leads to a proliferation of distinct configurations. We present an algorithm that partially restores permutation symmetry to such polydisperse packings.  
\end{abstract}

\maketitle  

A collection of $N$ identical soft spheres can be packed into a box in a multitude of stable configurations.  Let us count the ways.  We can do this in simulations on a computer. Different stable states can be accessed either by randomizing the initial particle positions or by imposing deformations that lead to particle rearrangements.   But, aside from simply cataloging the resulting states, can we predict how many configurations exist or find some order in this seeming chaos of a plethora of results?  At the outset, we confess ignorance!

In $d$-dimensions, the $Nd$-dimensional potential-energy landscape of the particle packing consists of basins whose minima each correspond to a single mechanically-stable configuration. The probability of being found when sampled randomly is proportional to the volume of a potential-energy basin; those with larger volume will be visited more often~\cite{o2002random,o2003jamming,xu2005random,doye2005characterizing,xu2011direct,frenkel2013other}.
As $N$, $d$, or other physical degrees of freedom such as particle shape or polydispersity increase, the number of distinct stable particle arrangements, $n_p$, grows rapidly. 

In general, it is considered rare to land in one configuration repeatedly unless the system is exceedingly small as was demonstrated in $d=2$ for $N \le 16$ bidisperse disks (a 50:50 mixture with diameter ratio $=1.4$) where Xu et al. were able to sample a significant fraction of the available stable configurations~\cite{xu2005random,xu2011direct}. They concluded that the distribution of basin volumes follows a log-normal distribution around an average basin size.  As $N$ increases, each basin occupies a smaller and smaller fraction of the entire configuration space so that it becomes more difficult to enumerate them all.  

\begin{figure}[h!]
\centering
\includegraphics[ width=6.5cm]{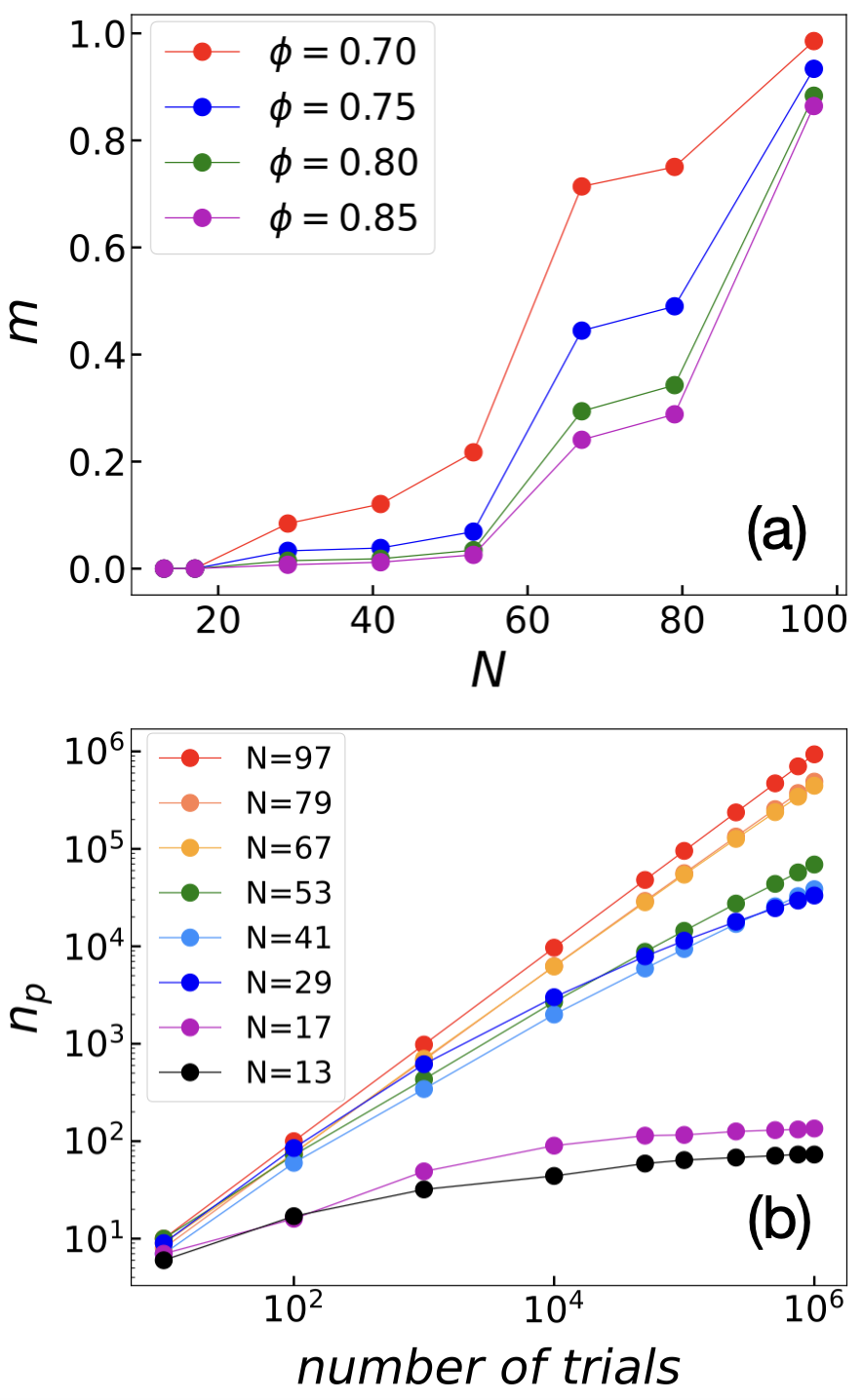}%
\caption{ (a) The fraction of distinct stable configurations, $m$, found in $10^6$ random samples of $d=3$ monodisperse packings versus system size, $N$. Below $N = 50$, relatively few stable states are found.  As $N$ increases, the fraction of new states found proliferates rapidly. At the four packing fractions, $\phi$, shown in the legend, $m$ decreases with increasing $\phi$.
(b) The number of distinct stable configurations, $n_p$, versus the number of trials (random initial conditions) at $\phi = 0.75$ for different $N$.  At small $N$, $n_p$ saturates while at large $N$, only a few states are found more than once.
}
\label{fig:mono}
\end{figure}

\begin{figure*}[t!]
\centering
\includegraphics[ width=18cm]{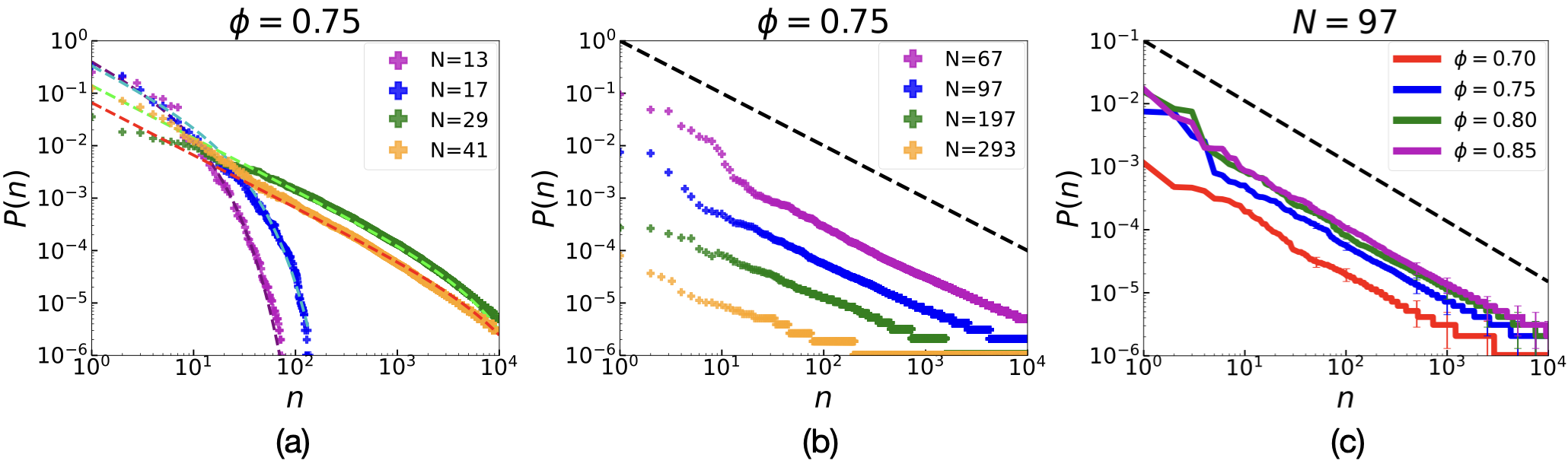}%
\caption{(a) The probability of finding a basin, $P(n)$, versus the rank order, $n$, at $\phi = 0.75$ for $N < 50$. (b) Same as (a) for larger systems: $N > 50$.  Even for $N = 293$, there are nearly $2$ decades of approximately power-law behavior. (c) $P(n)$ for packings with $N = 97$ at four packing fractions. The dashed lines in (a) are fits to: $P(n) = A_N\frac{1}{n}\myexp^{-n/n_0}$. The black dashed lines in (b,c), are guides to the eye with slope $-1$. In all panels, each data set is obtained from $10^6$ randomly sampled monodisperse packings.}
\label{fig:mono-scaling}
\end{figure*} 

In this paper, we analyze the distribution of stable configurations of soft \textit{monodisperse} spheres (in which all diameters are identical) in $d=3$; we sample configuration space by minimizing the system's potential energy after starting from randomly chosen initial particle positions.  For all system sizes studied, we were surprised to find that the distribution of basin volumes has an underlying order. We suggest that this order controls many of the static and dynamic properties of glassy systems by influencing how a system evolves under perturbations that destabilize the packing structure.

For small system sizes, $N$, we repeatedly find the same stable configurations.  However, the number of configurations is much smaller than was observed in the bidisperse $d=2$ systems because, for our monodisperse packings, the identical particles can be permuted without changing the packing.  The extra states added by permutations in a bidisperse (and more generally in a polydisperse) system overwhelm the number of intrinsic states of the monodisperse system~\cite{ozawa2018configurational}.  Here we introduce a protocol that allows permutation symmetry to be partially restored in a polydisperse system. This allows us to investigate the relationship between polydisperse and monodisperse packings.  
\begin{figure}[h!]
\centering
\includegraphics[ width=8.5cm]{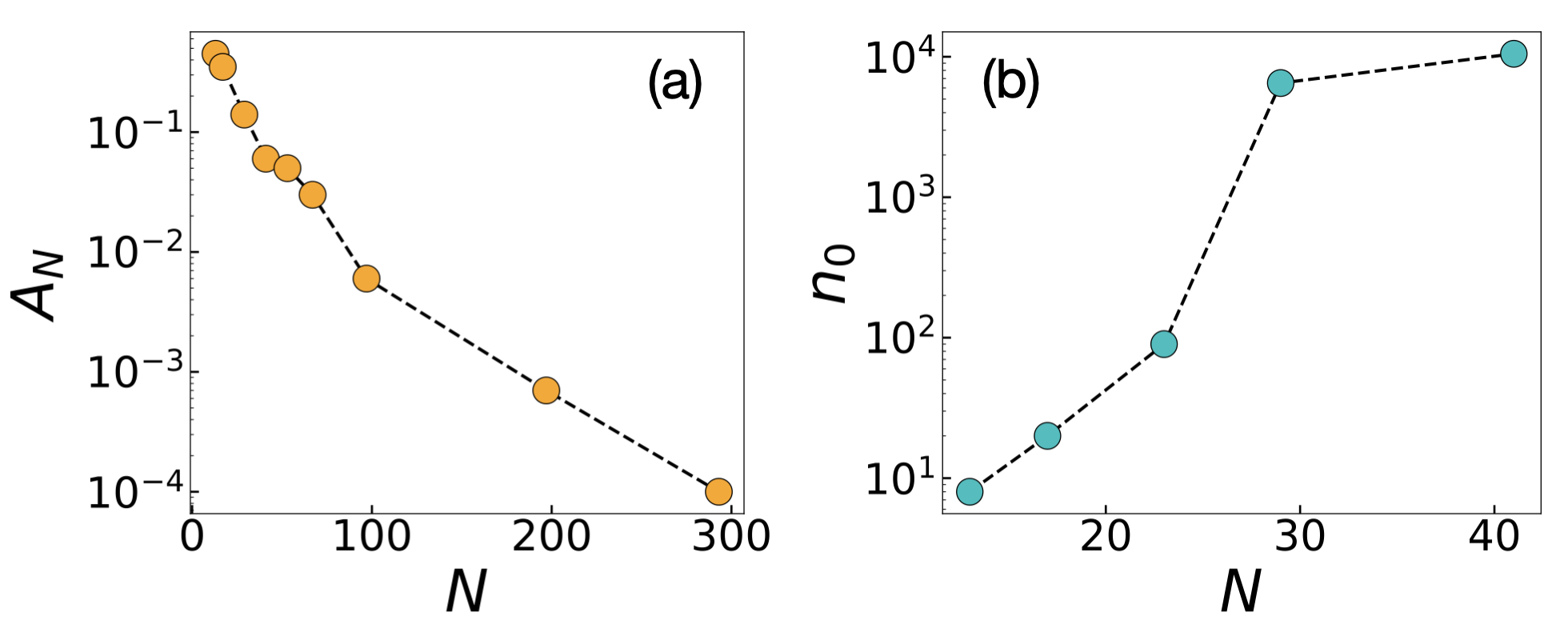}%
\caption{(a) The prefactor $A_N$ versus system size, $N$, for monodisperse packings at packing fraction $\phi = 0.75$. $A_N$ is obtained from a fit to Eq.~\ref{eq:n-scaling} with $\beta = 1.0$. (b) The cutoff value, $n_0$, versus $N$ for data in Fig.~\ref{fig:mono-scaling}a. }
\label{fig:mono-scaling-A-n0}
\end{figure}

\textit{The monodisperse landscape --  counting:} We first investigate monodisperse soft-sphere systems in a cubic box in $d=3$ with periodic boundary conditions.  We use harmonic interactions between particles $i$ and $j$ with radii $R_i$ and $R_j$ located at positions $\mathbf{r}_i$ and $\mathbf{r}_j$: 
\begin{equation}
    E_{i,j} = \ \sum_{i,j} \ \epsilon_0 \ \left( 1 - \frac{|\mathbf{r}_i - \mathbf{r}_j |}{R_i + R_j} \right)^{2} \Theta \left( 1 - \frac{|\mathbf{r}_i - \mathbf{r}_j |}{R_i + R_j} \right),
    \label{eq:energy}
\end{equation}
where $\epsilon_0$ is a constant setting the energy scale, $\Theta (x)$ is the Heaviside step function, and $R_i = R_j = R$ when the packings are monodisperse. We study packing fractions $0.70 < \phi < 0.85$ with the number of particles $13 < N < 293$. In all of our studies of the monodisperse landscape, we sample an ensemble of $10^6$ randomly chosen initial conditions (\textit{i.e.}, we choose random numbers for each coordinate of each particle). We minimize the energy of each configuration to quad precision using a GPU based implementation of the FIRE algorithm~\cite{morse2014geometric}. 

We determine whether two initial configurations end in the same state by comparing the energies of minimized states. 
If they are the same to $32$ decimal places, we then check to see if the particles have the same connections. We compare the connectivity of two packings using subgraph isomorphism if they are monodisperse~\cite{juttner2018vf2++}, and by relabeling particles based on their size if they are polydisperse. There are some symmetries, for example, reflections, rotations by $90^{\circ}$ in a square box, and rigid translations, that we \textit{do not} count as distinct.
  
Our results illustrated in Fig.~\ref{fig:mono}a show the fraction, $m$, of distinct stable configurations in an ensemble divided by the number of trials. For all densities studied, at $N=13$, the number of distinct configurations found when the landscape is randomly sampled $10^6$ times 
is $n_p \le 102$; on increasing the system size to $N=17$, this number grows to $n_p \le 220$.  
We note that the most frequently sampled basins are not necessarily the ones with the lowest potential energy.  Fig.~\ref{fig:mono}a shows a consistent trend: as $\phi$ increases the number of distinct configurations decreases. Closer to the jamming threshold there is less overlap between spheres and there is more room for stable rearrangements. This results in a more rugged and complex landscape~\cite{dennis2020jamming,artiaco2020exploratory}. 

As $N$ grows even larger, the number of distinct stable configurations, $n_p$, grows rapidly. Above $N=50$, $n_p$ rises sharply until it begins to approach $10^6$, resulting in $m \to 1$.   
In Fig.~\ref{fig:mono}b, we show $n_p$ as a function of sampling ensemble size at $\phi = 0.75$. For systems with $N \le 17$, the curves approach a plateau suggesting that we have effectively explored nearly all the available basins. For larger $N$, the curves show hardly any saturation as new, un-visited configurations continue to be found at nearly the same rate as they were initially.
These results, especially at large $N$, show that $10^6$ trials is not an exhaustive sampling of all the basins. The sheer enormity of the number of  basins results in only a minute fraction of configurations falling into the same basin multiple times~\cite{martiniani2016turning}.  

\textit{The monodisperse landscape -- statistical order:}  Fig.~\ref{fig:mono} shows that the basins must have very different volumes, $V$, because some configurations are found repeatedly while others are found only once or not at all. This was also noted in the $d=2$ studies of bidisperse systems~\cite{xu2005random,xu2011direct}. In order to explore this variation, we rank the basin volumes in descending order: $n=1$ is the largest basin which has been found the most times; $n=2$ is the second largest basin, \textit{etc.} Fig.~\ref{fig:mono-scaling} shows, for a system of size $N$, the probability, $P_N(n)$, that the $n$th basin would be found. $P_N(n)$ is the total number of times a basis was found divided by the number of trials. By construction, $P_N(n)$ must monotonically decrease and its sum over all $n$ must be unity.  For clarity, we split the results into two bins: Fig.~\ref{fig:mono-scaling}a, shows the results for $N<50$ and Fig.~\ref{fig:mono-scaling}b shows the results for $N>50$.

For the larger system sizes ($N>50$) shown in Fig.~\ref{fig:mono-scaling}b, $P_N(n)$ appears to approach a simple scaling behavior: $P_N(n) \propto n^{-\beta}$ with $\beta \approx 1.0 \pm 0.1$. We note that there are detectable deviations from the straight-line behavior on these graphs. Nevertheless, the overall trends are clear. For large $n$, the trend is cut off by the number of samples in our ensemble, \textit{i.e.}, $10^6$.  This behavior persists at least out to $N=293$, the largest system size we investigated. In Fig.~\ref{fig:mono-scaling}c, we show the data as a function of packing fraction, $\phi$, for a single system size, $N=97$.  The variation with $\phi$ does not alter the scaling behavior and only shifts the curves vertically.

In the smaller systems ($N<50$) shown in Fig.~\ref{fig:mono-scaling}a, this scaling is truncated. We therefore fit each dataset by a power-law that is cut off by an exponential factor: 
\begin{equation}\label{eq:n-scaling}
P_N(n) = A_N n^{-\beta} \myexp^{-n/n_0},
\end{equation}
where $A_N$ and and $n_0$ are a prefactor and a cutoff that depend only on system size, $N$.  For the fitting we have chosen $\beta = 1.0$. $A_N$ versus $N$ is shown in Fig.~\ref{fig:mono-scaling-A-n0}a. In Fig.~\ref{fig:mono-scaling-A-n0}b we show $n_0$ versus $N$ for $N<50$. For $N>50$, the value of $n_0$ becomes too large to be extracted directly and $P_N(n) = A_N n^{-\beta}$ is a good description of the data over the available range.    While our data show that $\beta \approx 1.0 \pm 0.1$, we are unable to determine clearly if there are slight variations or a drift in its value as a function of $N$ or $\phi$.

Theses results were for monodisperse systems in $d=3$.  In this case with permutation symmetry (\textit{i.e.}, that particles can be permuted in any order -- amounting only to relabeling the particles -- without affecting the system), the number of distinct states is far less than we naively expected.  In this case, the underlying structure of the packings was made manifest so that surprising statistical order emerged.

\textit{Breaking -- and partial restoration of -- permutation symmetry:} Breaking permutation symmetry by introducing small amounts of polydispersity results in an enormous increase in the number of distinct minima in the energy landscape because there are $N!$ ways of rearranging the particles in each basin.  When the particles are not identical, these configurations can be distinguished from each other.  Even for $N=13$, any polydispersity makes it impossible for us to access the same state twice by random sampling initial conditions.

\begin{figure}[h!]
\centering
\includegraphics[ width=8cm]{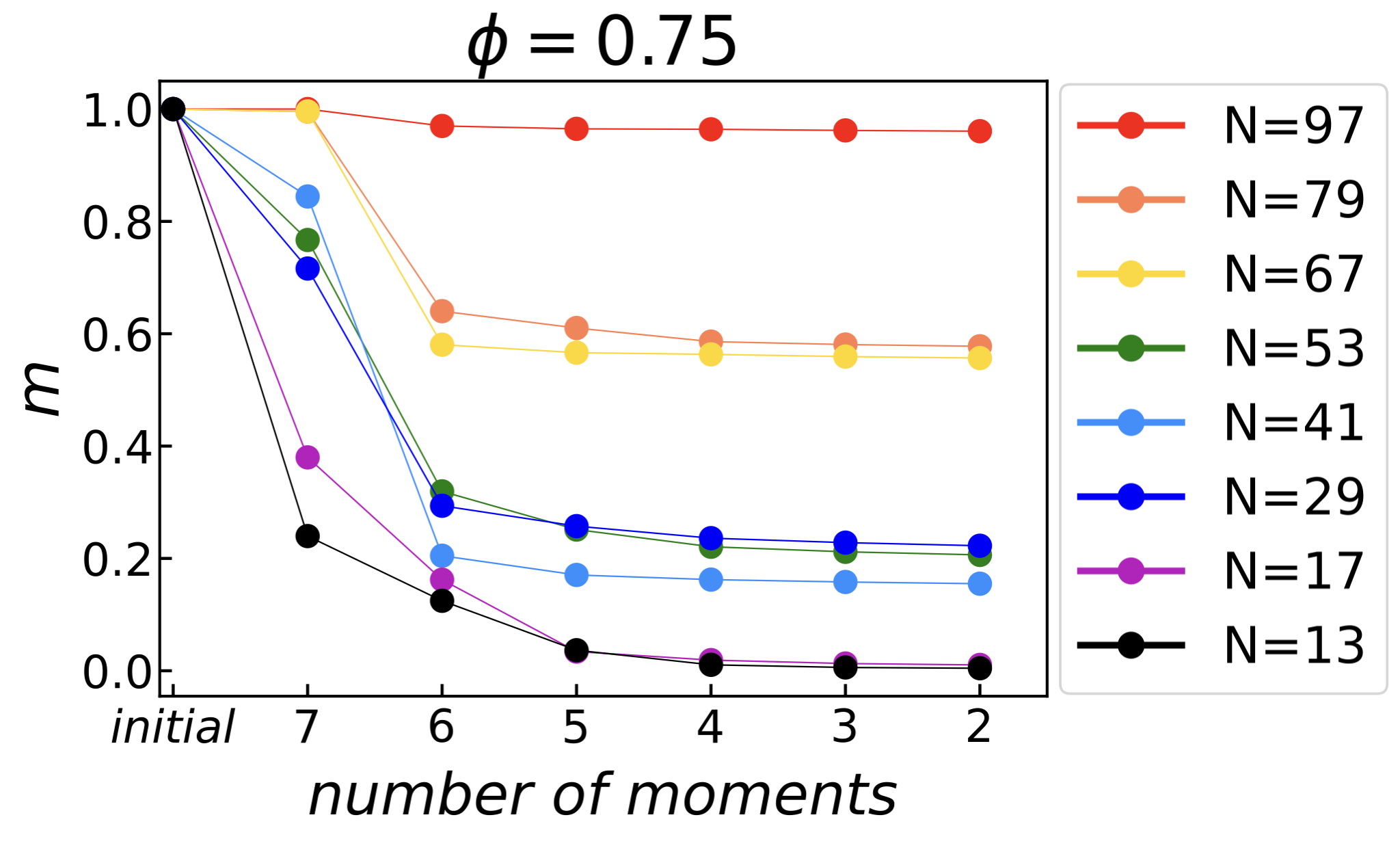}%
\caption{ The fraction of distinct stable configurations, $m$, in $10^4$ random samples at packing fraction $\phi=0.75$ and polydispersity $\sigma = 0.01$ versus the number of moments of radii that are constrained during minimization. As fewer moments are constrained, the number of distinct stable configurations drops dramatically. $m$ typically grows with $N$. }
\label{fig:moments}
\end{figure} 
If the degree of polydispersity is small, a single large basin in the monodisperse landscape will be replaced by a set of smaller ones -- one for each permutation of the particles. Some traits of the large monodisperse basin can still be detected in the individual offspring. Thus, despite the enormous effect of polydispersity on breaking the permutation symmetry, the addition of very small amounts of polydispersity can nevertheless be viewed as a perturbation to monodisperse packings.

In order to see this, we apply the following algorithm to partially ``restore'' the permutation symmetry.  We create an ensemble of $N$ polydisperse spheres; each sphere has a different radius chosen from a log-normal distribution of particle sizes of width $\sigma_R$ and mean $\braket{R}$. We define polydispersity as $\sigma  = \frac{\sigma_R}{\braket{R}}$.  Once these radii have been chosen, we create an ensemble using the \textit{same set of radii} for each random configuration.

To minimize the  energy, in addition to moving the particle positions we also allow the particle radii to change~\cite{hagh2022transient}. This process is reminiscent of the swap Monte Carlo~\cite{ninarello2017models,ikeda2017mean, brito2018theory, berthier2019efficient}, but instead of doing individual swaps between particles of different sizes, we let all radii adjust simultaneously (more similar to collective swap algorithms~\cite{ ghimenti2024irreversible,kim2024towards}. In this process, the radii are considered as degrees of freedom. However, unrestricted changes in radii can potentially cause some radii to go to zero and un-jam the system. To circumvent this, we constrain certain moments of the distribution by removing components of radii forces, $\frac{\partial E}{\partial R_i}$, perpendicular to the $\sum_{i} R_{i}^{\alpha} = c $ plane where $c$ is a constant. Initially, we fix seven moments $\alpha = (-6, -3, -1, 1, 2, 3, 6)$, maintaining the radii distribution close to the original while enhancing stability~\cite{hagh2022transient}. The choice of seven initial moments is arbitrary, as it only should ensure stability of packings without any un-jamming.
\begin{figure}[h!]
\centering
\includegraphics[ width=6.5cm]{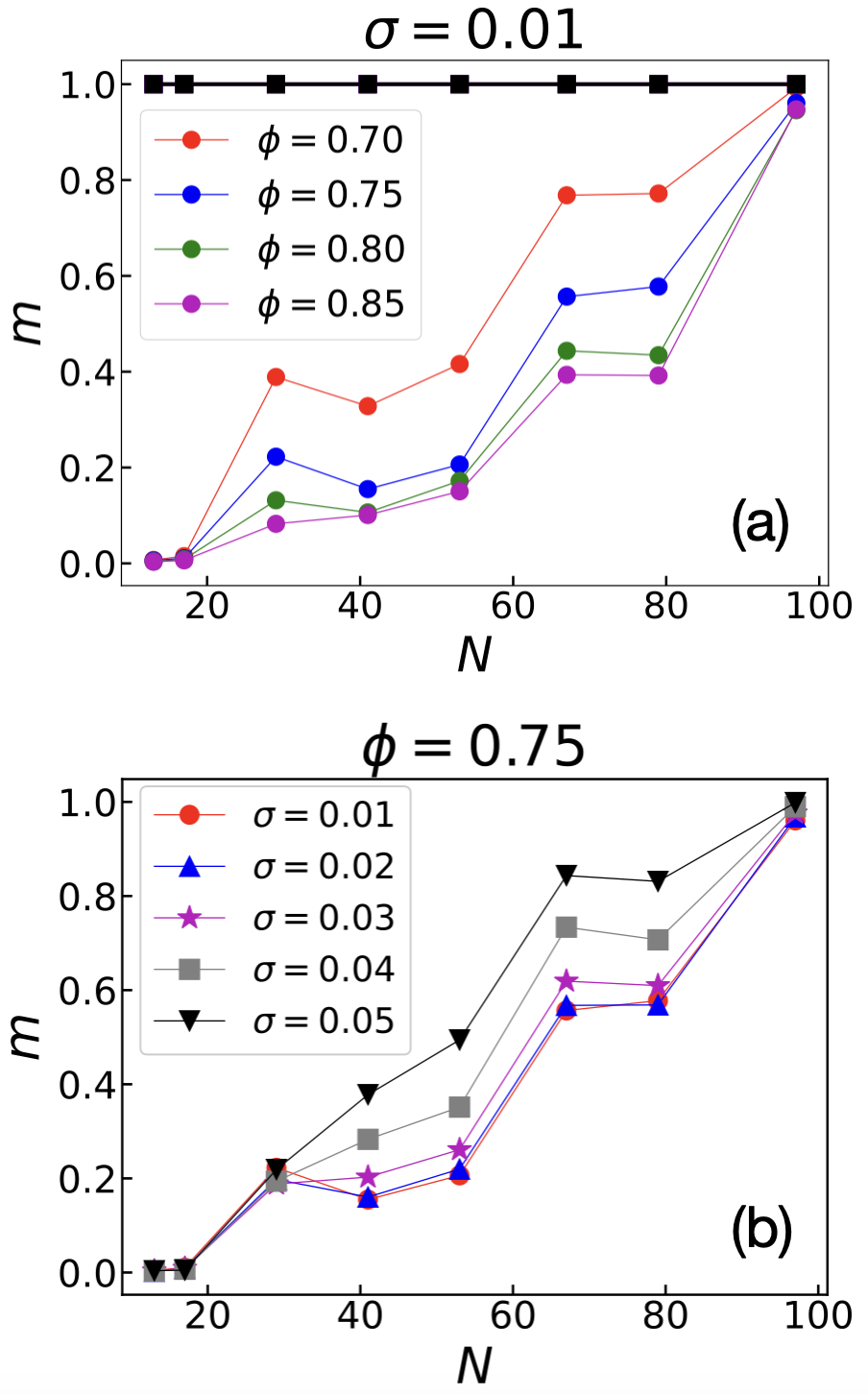}%
\caption{(a) $m$, the fraction of distinct configurations with polydispersity $\sigma = 0.01$ versus system size, $N$, at different packing fractions, $\phi$.  Circles show that after application of the minimization-and-replacement algorithm, many configurations are found multiple times.  Squares (at $m=1$) show that no repeated states are found before application of the algorithm. 
(b) $m$ versus $N$ at different values of $\sigma$ at $\phi = 0.75$ after applying the algorithm. In both panels, each point is determined from an ensemble of $10^4$ trials.}
\label{fig:poly}
\end{figure}  

In the final stage of the protocol, we replace the output of the radii minimization with the \textit{original} particle radii.  We do this based on their size ranking: the largest particle in the radii-minimized configuration is replaced with the largest particle size in the original packing, then we do the same for the next largest particle \textit{etc.} until we have replaced all the particles.  Finally, we minimize the energy one more time without allowing the radii to change. 
This procedure guarantees a stable configuration with the identical radii as the original distribution.
By going through this cycle of (i) energy minimization with respect to radii and positions, (ii) radii replacement, and then (iii) energy minimization only with respect to positions, we obtain a decrease in the number of distinct packings as many allowed packings in the original energy landscape are congregated into ultrastable configurations.  

We can further improve on this if we repeat the minimization, this time constraining only six out of the initial seven moments of the radii distribution. The process continues, progressively reducing the number of fixed moments during radii minimization and replacing the resulting radii with equivalent values from the original packing, until we constrain only two moments, $\alpha = \{ -3, 3\}$, during minimization. Two is the minimum number of constraints required to keep all packings jammed across various sizes and packing fractions. The gradual reduction in the number of constrained moments makes the minimization procedure more gentle so that the packing does not lose stability.  

This process vastly reduces the number of distinct packings as shown in Fig.~\ref{fig:moments}.  Whereas the original polydisperse sampling never recovered the same configuration twice, the samples formed after minimization with respect to radii and replacement by the original sizes funnels many states into the same deep minimum as shown in Figs.~\ref{fig:poly}a,b for various packing fractions, $\phi$, and polydispersities, $\sigma$. The fraction $m$ after radii minimization and replacement is small across range of polydispersities $\sigma \le 5 \%$.  It becomes less pronounced for larger  $\sigma$ in larger system sizes.

\begin{figure*}[t!]
\centering
\includegraphics[ width=18cm]{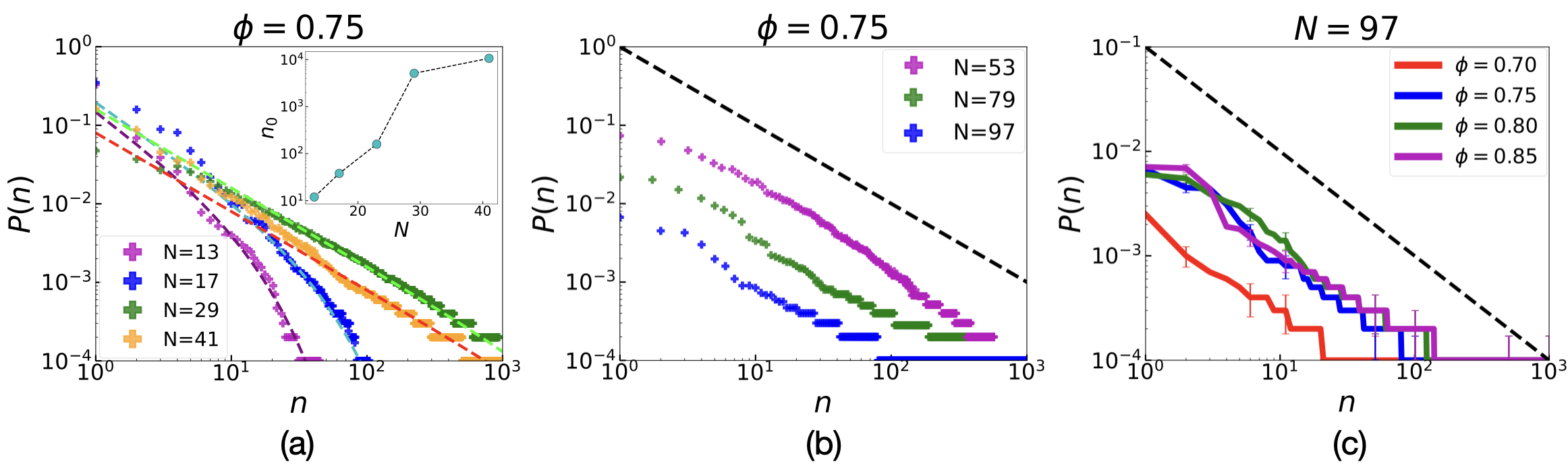}%
\caption{$P(n)$, the probability of finding a basin versus $n$, the rank ordering of basin sizes, in polydisperse packings after having been processed through the algorithm with two constrained moments of the distribution. (a) $P(n)$ for $N < 50$ at $\phi = 0.75$. The dashed lines are fits to: $P(n) = A_N\frac{1}{n}\myexp^{-n/n_0}$. . The inset shows $n_0$ grows rapidly with $N$. (b) $P(n)$ for packings with $N > 50$, at $\phi = 0.75$. (c) The probability for packings with $N = 97$ particles at four values of $\phi$. The black dashed lines in (b,c), are guides to the eye with slope $-1$. In all panels, each data set is obtained from $10^4$ randomly sampled polydisperse packings.
}
\label{fig:poly-scaling}
\end{figure*}

To emphasize the relationship of these states in the polydisperse and monodisperse systems, we show that they conform to the same statistical regularity as the monodisperse packings. As illustrated in Fig.~\ref{fig:poly-scaling}, the probability scales inversely with rank $n$, as found in Fig.~\ref{fig:mono-scaling}.

\begin{figure}[h!]
\centering
\includegraphics[ width=6.5cm]{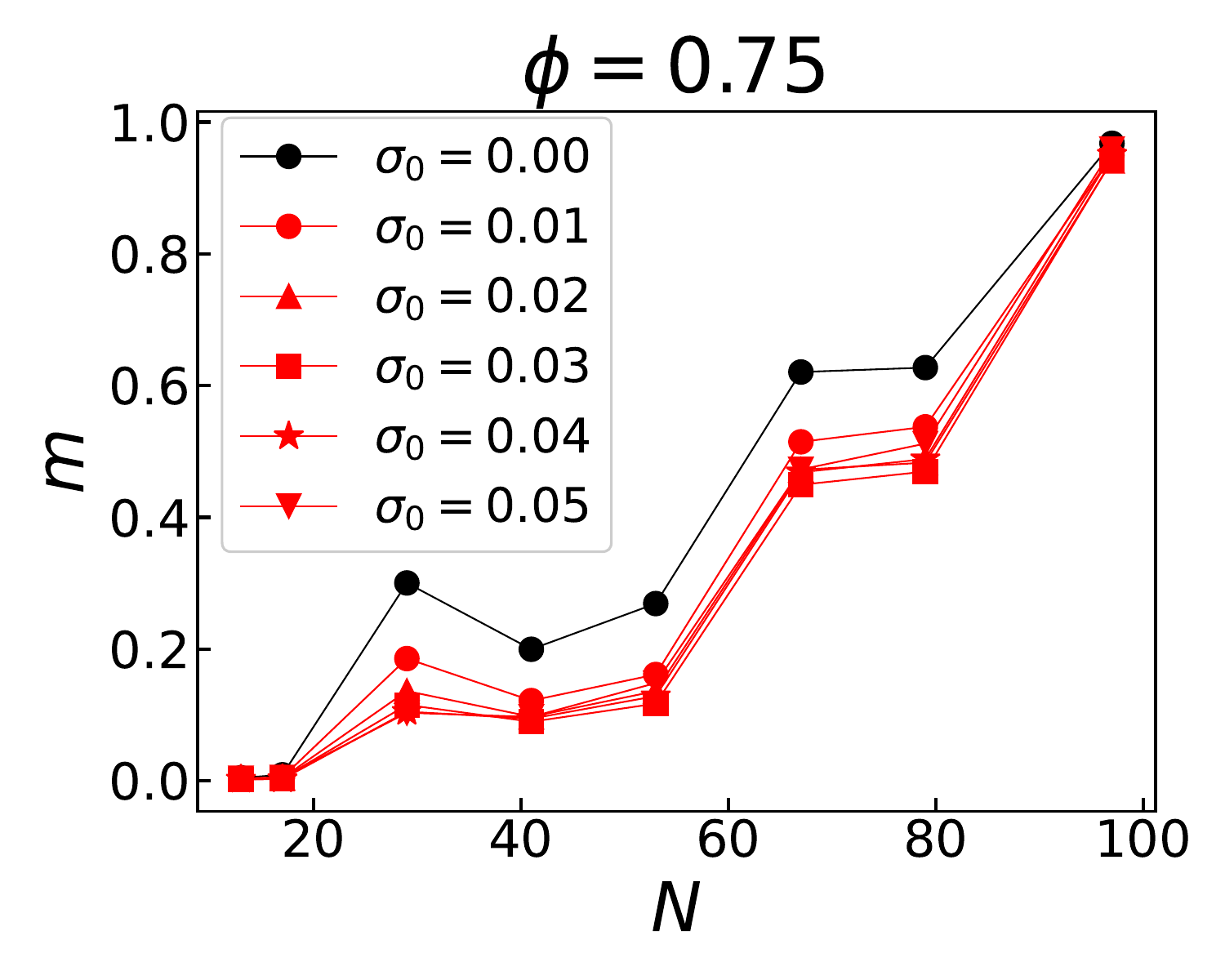}%
\caption{ The fraction of distinct stable configurations, $m$, versus system size in random monodisperse packings (black) and in polydisperse packings (red) produced by replacing all radii at initial polydispersity $\sigma_0$ with a single value such that $\phi = 0.75$. Each data point is from $10^{4}$ trials.  The polydisperse packings are those shown in Fig.~\ref{fig:poly}b
}
\label{fig:monoFrom2m}
\end{figure}

As a final demonstration that these repeated minima are related to the large basins on the monodisperse landscape, we set all the radii of the minimized polydisperse packings to the single monodisperse value while fixing the packing fraction. When we equilibrate them with respect to positional degrees of freedom only, we find that all the packings that clustered into a single basin in the polydisperse landscape end up in one of the previously observed larger basins on the monodisperse landscape. This holds true for all studied polydispersies as shown in Fig.~\ref{fig:monoFrom2m}. 
The black data show the equivalent fractions of random monodisperse packings. This demonstrates that our algorithm partially recovers the permutation symmetry in the polydisperse landscape. 

\textit{Discussion:} For monodisperse systems, we find an  inverse relationship between a basin's volume-based rank order and the volume of that particular basin.  While the underlying cause of this relation remains unexplained, we nevertheless suggest that this form of order must have significant consequences for the properties of glasses and jammed solids. When randomly sampled, the probability of finding the system in any configuration with volume between $V$ and $V+dV$ is the weighted average $VP(V)dV$.  
Given $P(n)\propto V \propto n^{-\beta}$ for $n<n_0$ with exponent $\beta = 1.0$, we find $P(V) = P(n)|\frac{dn}{dV}| \propto P(n) n^2 \propto n$ so that $VP(V) \approx$ constant.
This is a flat probability for all volumes; the system will be in a basin which is equally likely to have a large or small volume.  
Thus, when a stable configuration is perturbed, there is a likelihood that it will eventually fall into a larger basin; shearing a random packing can continually drive the system into these larger wells, collecting many small basins into progressively larger ones.  

We note that using Eq.~\ref{eq:n-scaling} we can answer the question posed in our opening paragraph and estimate $n_0$, which is a good approximation for the total number of configurations for a system of size $N$. The integral of $P_N(n)$ over all $n$ must be unity. If the exponent $\beta=1.0$ we find that the cutoff is given by $n_0 = \myexp^{1/A_N}$. $A_N$ is shown in Fig.~\ref{fig:mono-scaling-A-n0} for the monodisperse data at $\phi=0.75$. We note that the entire distribution is not necessary to determine $A_N$ (and therefore $n_0$) since it can be estimated from $P_N(n)$ at small $n$.

It is natural to think of our algorithm as a process for partially restoring permutation symmetry to a polydisperse packing. The minimization with respect to radii and eventual replacement with the original distribution  allows the radii of all the particles in the system collectively to find a particularly favorable ground state. The process erases the distinctiveness that was created by the initial choice of individual particle radii; the original labeling of the particles (representing slightly different radii) can be interchanged freely -- as if permutation was allowed -- to arrive at the optimal minimized configurations.
This procedure not only identifies 
minima in the polydisperse landscape which are ultra-stable, but also shows that these states transform into the \textit{same} states as were found in the monodisperse packing. In addition, the statistics of these minima recover the ordered patterns present in the monodisperse system. These results all suggest that the protocol provides a way to partially restore permutation symmetry in the polydisperse landscape.

Generally, breaking physical symmetries 
introduces extra possible distinct states, rendering the landscape more rugged. The present study demonstrates that introducing and manipulating radii degrees of freedom systematically identifies a subset of states that reveal scars in the polydisperse landscape created by the permutation symmetry of the monodisperse situation. This is important in its own right because it indicates how various powerful methods, such as swap Monte-Carlo~\cite{ninarello2017models,ikeda2017mean, brito2018theory, berthier2019efficient}, based on the swapping of different-size particles, can be related to the physics of monodisperse amorphous solids where swapping particles no longer changes the landscape.

A number of questions present themselves.  Chief among them are: (i) What is the cause of the ordering that we find in the distribution of basin volumes?  This is a purely geometric question involving how distinct sphere packings fill up configuration space. (ii)  How does this underlying order in the distribution of packing volumes affect the physical properties of disordered solids?  This is a physical question about how dynamics, under equilibrium conditions, or external perturbations drive the evolution of a system to visit more or less probable basins in the vast energy landscape.  Knowing that the distribution of basin volumes has order, clearly influences how one should attack such a problem. It entails understanding how different basins are related to one another in configuration space. (iii)  What are the implications of $VP(V) \approx$ constant that was derived from the particular form of the volume distribution that we discovered?  This suggests a statistically maximally heterogeneous underlying landscape where ending up in basins of any size are all equally likely.  (iv)  Does this result, found for three dimensions, have counterparts in other dimensions?  If so, one might gain analytic traction by going to the mean-field (\textit{i.e.}, $d=\infty$) limit.  (v)  Does the echo and retrieval of the monodisperse states in the polydisperse landscape indicate that the correlations uncovered in this work are present in the polydisperse situation as well? The role of permutation symmetry breaking is clearly important since in polydisperse packings the $N!$ multiplication for the number of minima dwarfs the number of states in the monodisperse system.  (vi) Returning to where this inquiry started, how are memories preserved in disordered packings? We believe the results presented in this study provide some insight into how this can occur.

\textcolor{white}{whitespace}

We thank Eric Corwin, Andrea Liu, Thomas Witten, and Francesco Zamponi for discussions. SRN was supported by the University of Chicago Materials Research Science and Engineering Center, NSF-MRSEC program under award NSF-DMR 2011854.  VFH was supported by the US Department of Energy, Office of Science, Basic Energy Sciences, under Grant DE-SC0020972. 
\bibliography{main}
\end{document}